\documentclass[pdflatex,sn-basic,Numbered]{sn-jnl}

\usepackage{graphicx}%
\usepackage{amsmath,amssymb,amsfonts}%
\usepackage[title]{appendix}%
\usepackage{textcomp}%
\usepackage{booktabs}%
\usepackage{tabularx}
\usepackage[utf8]{inputenc}
\usepackage{microtype}
\usepackage{soul}
\usepackage[dvipsnames]{xcolor}
\usepackage[normalem]{ulem}
\usepackage{tikz}
\usetikzlibrary{positioning,arrows.meta}
\newcolumntype{R}[1]{>{\hsize=#1\hsize\raggedright\arraybackslash}X}
\newcolumntype{C}{>{\centering\arraybackslash}X}
\newcolumntype{Z}[1]{>{\hsize=#1\hsize\centering\arraybackslash}X}
\newcommand{\runinhead}[1]{%
  \par\addvspace{0.6\baselineskip}%
  \noindent\textbf{#1}\ \ignorespaces}

\begin{document}

\title[The Ethics of Artificial Intelligence in Military Operations]{The Ethics of Artificial Intelligence in Military Operations}

\author*[1,2]{\fnm{Nicolas} \sur{Drapier}}\email{nicolas.drapier@sas-impact.fr}
\author[2]{\fnm{Florian} \sur{Mauberger}}\email{florian.mauberger@sas-impact.fr}
\author[1]{\fnm{Aladine} \sur{Chetouani}}\email{aladine.chetouani@univ-paris13.fr}
\author[2]{\fnm{Aurélien} \sur{Chateigner}}\email{aurelien.chateigner@sas-impact.fr}
\affil*[1]{\orgdiv{L2TI Laboratory}, \orgname{Université Sorbonne Paris Nord}, \orgaddress{\street{99 Avenue Jean-Baptiste Clément}, \city{Villetaneuse}, \postcode{93430}, \country{France}}}
\affil[2]{\orgname{SAS Impact}, \orgaddress{\street{1 Rue Sainte-Anne}, \city{Orléans}, \postcode{45000}, \country{France}}}

\abstract{Deep learning systems now mediate military decisions to use force, yet their internal logic resists inspection, their evaluation practices are gameable, and their deployment fractures accountability across dispersed stakeholders. The ethical challenge posed by these systems is fundamentally epistemic: not just whether autonomous weapons should be permitted to kill, but whether the conditions for responsible human judgment can survive when critical functions are delegated to opaque algorithms.

We show that this epistemic condition produces a concrete accountability gap: responsibility diffuses across designers, operators, and policymakers while International Humanitarian Law presupposes capacities for judgment that current AI systems lack. To address this gap, we propose a governance framework that proceduralizes ethical constraints through named accountability roles, adversarial auditing with undisclosed benchmarks, tiered deployment thresholds, and a proposed NATO evaluation standard.

Counterfactual analysis of eight documented cases (1988-2025) shows that each governance mechanism addresses a documented class of failure, but no single safeguard suffices in isolation: effective governance of military AI requires not only technical constraints but the institutional infrastructure to keep human judgment meaningful.}


\maketitle

% ==========================================
%               INTRODUCTION
% ==========================================
\section{Introduction}
\label{sec:introduction}
The ethical debate surrounding military AI has largely centered on a binary question: should autonomous systems be permitted to select and engage targets? This framing has motivated important policy initiatives, including the open letter by AI researchers calling for a ban on offensive autonomous weapons~\cite{fli2015openletter} and the ongoing deliberations within the UN Convention on Certain Conventional Weapons. Yet it obscures a more fundamental problem. The challenge is not only whether machines should be allowed to kill, but whether the epistemic conditions for responsible decision-making can be preserved when critical functions are delegated to systems whose internal logic resists inspection.

These systems are already operational. Targeting, surveillance, logistics, and command support increasingly depend on deep learning operating at speeds and scales beyond human cognitive capacity. We call this shift \emph{Algorithmic Warfare}: the integration of autonomous inference into the chain of command, from sensor processing to engagement decisions.

This paper argues that the ethical crisis of military AI is, at root, an epistemic crisis. The technologies driving this transformation range from autonomous targeting to predictive logistics~\cite{Scharre2018}, and their proliferation is reshaping strategic stability~\cite{Horowitz2019}. Deep learning models are opaque in ways structurally distinct from prior military technologies~\cite{doi:10.1177/2053951715622512, Lipton2018-dy}: their decision processes elude human comprehension even when source code is available, and post hoc explainability techniques offer approximations that do not restore access to the model's reasoning~\cite{Yang2023-we, whydoexplanationsfail2024}. This opacity compounds across the chain of command: it distorts operator trust~\cite{Goddard2012-cc, Lyell2017-at}, fragments accountability across dispersed stakeholders~\cite{Nissenbaum1996-NISAIA, Matthias2004, Sparrow2007}, and strains the applicability of International Humanitarian Law, whose core principles presuppose capacities for judgment that current AI systems lack~\cite{sharkey1008, Heyns2013, boulanin2020autonomous, Bhuta2016}. Responsibility does not vanish when decisions are delegated to algorithms. It is displaced, obscured, and redistributed through institutional mechanisms that simulate accountability without delivering it.

Our central claim is that governable military AI requires epistemic infrastructure: institutions, procedures, and technical constraints designed to preserve the conditions for human judgment under irreducible uncertainty. This is not a call to halt military AI adoption, which would be neither realistic nor strategically responsible, but to ensure that deployment proceeds within governance structures adequate to the technology's demands.
This paper makes two contributions:
\begin{enumerate}
    \item A diagnostic framework showing how the structural opacity of deep learning interacts with adversarial fragility, benchmark gaming, and defense-contracting incentives to produce three compounding epistemic failures (an \emph{illusion of understanding}, an \emph{illusion of accuracy}, and an \emph{illusion of determinism}) that erode both accountability and the practical applicability of International Humanitarian Law (\ref{sec:part1}).
    \item A procedural governance framework (named accountability roles, adversarial auditing with undisclosed benchmarks, a proposed NATO evaluation standard, tiered deployment thresholds, and interface design principles) evaluated through counterfactual analysis of eight operational cases spanning 1988-2025 (\ref{sec:design_principles}).
\end{enumerate}
% ==========================================
%               INTRODUCTION
% ==========================================

% ==========================================
%                BACKGROUND
% ==========================================
\section{Background}
\label{sec:part1}

\subsection{Opacity, Explainability, and Adversarial Fragility}
\label{ssec:opacity}

The three sources of algorithmic opacity identified by Burrell~\cite{doi:10.1177/2053951715622512} converge simultaneously in military AI: proprietary algorithms and classification regimes enforce deliberate dissimulation, rapid development cycles produce organizational ignorance, and deep neural networks are mathematically opaque by construction~\cite{Lipton2018-dy}. Post hoc explainability methods such as LIME~\cite{10.1145/2939672.2939778}, SHAP~\cite{10.5555/3295222.3295230}, and Grad-CAM~\cite{Selvaraju2017-uk} can aid debugging~\cite{Van_Zyl2024-tg}, but these are approximations that can be unfaithful to the model's actual decision process~\cite{Yang2023-we, whydoexplanationsfail2024}, and explanation quality fails to predict human-AI team performance~\cite{10.1145/3377325.3377498}. In military applications the deeper risk is what we term an \emph{illusion of understanding}. An analyst viewing a heatmap concentrated on a vehicle's turret may conclude the model identified a tank by its weapon system, while the model may be keying on background terrain or sensor artifacts. The explanation satisfies the cognitive need for justification without providing epistemic access. Partial transparency is therefore more dangerous than acknowledged opacity: complete opacity preserves skepticism, while the illusion of understanding actively suppresses it.

Adversarial fragility compounds this danger. Physically realizable perturbations systematically fool classifiers~\cite{brown2018adversarialpatch, 8578273}, defenses remain unstable~\cite{Carlini2017, Madry2018}, and models that achieve high accuracy on curated benchmarks degrade unpredictably under distributional shifts characteristic of operational environments~\cite{Hendrycks2019}. When such systems output predictions without calibrated uncertainty, the result is an \emph{illusion of accuracy}: the model reports high confidence on every classification, whether processing a clear image or a degraded sensor feed. Failures present identically to successes, and the error becomes visible only when its consequences materialize.

These two bodies of work are typically treated in isolation. We argue that their intersection is where the deepest risk lies: opacity prevents operators from detecting when a system crosses its competence boundary, and adversarial fragility means such crossings can be induced deliberately. No prior work examines this interaction as a unified epistemic condition specific to military decision-making.

\subsection{Trust, Institutional Incentives, and Systemic Vulnerabilities}
\label{ssec:trust}

The illusions of understanding and accuracy interact with institutional dynamics to produce a third: the \emph{illusion of determinism}. Clean dashboards and categorical outputs erase the probabilistic nature of the underlying computation. A commander who has observed correct identifications across hundreds of training exercises generalizes to the expectation that the system will perform identically in theater, where unfamiliar terrain, degraded sensors, and adversarial countermeasures invalidate that expectation. The categorical format provides no warning that the system has crossed the boundary of its competence. Operators are known to defer to automated recommendations, a disposition documented as automation bias~\cite{Goddard2012-cc, parasuraman1997humans, Cummings2004}. In these systems the problem starts earlier, before any deference: a categorical output reports a verdict without the confidence signal that would let an operator judge whether to trust it or not. Interface design mediates whether system limitations become perceptible~\cite{norman1999affordance}. Lee and See~\cite{lee2004trust} show that overtrust, calibrated trust, and distrust produce distinct failure signatures, while Risko and Gilbert~\cite{risko2016cognitive} characterize the mechanism as selective cognitive offloading.

The benchmarking ecosystem reinforces this dynamic. When evaluation criteria are known in advance, developers optimize for those criteria at the expense of broader robustness~\cite{Amodei2016ConcretePI, Russell2019}, and benchmark performance becomes a proxy for institutional confidence in systems that may fail under conditions never tested. Weight poisoning~\cite{kurita2020weightpoisoningattackspretrained, li2021backdoorattackspretrainedmodels} represents a further threat: an adversary can induce targeted failures that standard evaluations do not detect, and the poisoned model continues to perform well on benchmarks while behaving anomalously in high-stakes scenarios.

The growing dependence on private technology firms creates a parallel vulnerability. Project Maven~\cite{DoD2017Maven, DoDIG2022Maven} exposed this concretely: Google engineers who raised ethical concerns lacked formal channels to influence deployment, while DoD officials who authorized it lacked access to the underlying code or training data. The state owned the system, but the knowledge resided elsewhere. As Raji \textit{et al.}~\cite{Raji2023-jn} document, real-world AI failures in high-stakes domains typically stem from breakdowns in engineering practice. Military contexts, characterized by secrecy and operational pressure, amplify these gaps. Trust calibration and systemic vulnerability literatures both miss the institutional machinery that compounds these problems: benchmark gaming, contracting structures that separate knowledge from authority, and fragmented expertise that prevents any single actor from recognizing failure conditions.

\subsection{The Accountability Gap}
\label{ssec:accountability_gap}

AI-mediated decision-making dissolves the classical military chain of command into Nissenbaum's~\cite{Nissenbaum1996-NISAIA} ``many hands problem.'' The model is designed by engineers who may never witness its deployment, trained on datasets curated by scientists who will never select a target, operated by soldiers who cannot read its internal logic, and authorized by policymakers who lack both technical and operational knowledge. Each actor holds a fragment of agency; none fully causes the system's decisions. When failure occurs, responsibility is not discovered but negotiated~\cite{Nissenbaum1996-NISAIA}, and scapegoating becomes a structurally likely outcome. Matthias~\cite{Matthias2004} introduced the ``responsibility gap'' for learning automata; Sparrow~\cite{Sparrow2007} extended it to lethal autonomous weapons; and machines lack the intentionality that grounds moral accountability~\cite{Floridi2004-mu, Asaro2012}. If they cannot bear responsibility and human beings are too fragmented to deal with it on their own, existing structures provide no mechanism for attribution.

The concept of ``meaningful human control''~\cite{Santoni_de_Sio2018-tr} is intended to fill this gap, but it lacks operational precision. For oversight to be functionally effective, two conditions must hold: temporal authority (the power to veto, pause, or require stage-gated approval) and interpretive access (the ability to understand \emph{why} a system produced a given output, not just whether to endorse it). The international NGO Article 36~\cite{killingbymachine} has interpreted the obligations of states deploying autonomous weapons along these lines:
\begin{itemize}
    \item States must explicitly affirm that meaningful human control is required over individual attacks.
    \item Weapon systems operating without such control must be prohibited.
    \item States must publicly explain how they apply control over existing systems and justify why they consider them acceptable and lawful.
\end{itemize}

In practice, however, the system's output arrives with an aura of authority, and what begins as collaboration slides into deference. On Fischer and Ravizza's account of guidance control~\cite{Fischer1998-FISRAC-3}, the human remains responsible so long as the decision reflects their own critical judgment. But if the operator consistently aligns with the AI's output without challenge, their role becomes performative: they are not exercising judgment but affixing a stamp of approval. This is not real control. It is \emph{responsibility laundering}. Automation bias reinforces the dynamic~\cite{Goddard2012-cc, Lyell2017-at}, and military operators are rarely trained to recognize overconfidence, distributional shift, or adversarial manipulation~\cite{Strauch2017-hi}. The challenge is not to preserve human involvement as such, but to ensure that it is cognitively robust: the capacity to understand, interrogate, and override.

\subsection{IHL Under Strain}
\label{ssec:dih}

International Humanitarian Law rests on principles that presume human judgment, contextual awareness, and moral intentionality~\cite{Bhuta2016}, but the specific points of failure differ. Sharkey~\cite{sharkey1008} and Roff~\cite{Roff2014} show that distinction and proportionality require contextual interpretation that probabilistic classifiers cannot replicate: the difference between a combatant and a civilian holding a similar object is not a feature-space boundary but a moral assessment. Heyns~\cite{Heyns2013} adds that even if targeting were technically accurate, the physical and moral distance introduced by autonomy undermines the attribution on which legal accountability depends. Liu~\cite{Liu2012-jl} extends the analysis to situations current systems handle worst: recognizing surrender, medical evacuation, or \textit{hors de combat} status, where the legally required response is restraint, not classification.

Boulanin \textit{et al.}~\cite{boulanin2020autonomous} translate these requirements into three testable conditions for lawful deployment: foresight, administration, and traceability. Crootof~\cite{Crootof2022} reframes the problem. If AI-mediated outcomes cannot be fully predicted, accountability grounded in intent becomes unworkable. Her ``war torts'' model shifts the focus from the decision-maker's intent to the question of whether the decision-making process adhered to verifiable standards of care. This move from intent to procedural integrity is the conceptual foundation of our framework: it transforms an intractable philosophical question (who holds moral responsibility for an algorithmic output?) into a verifiable institutional one (what process was followed, and did it meet documented standards?). Yet the gap between this insight and operational practice remains wide. Jobin \textit{et al.}~\cite{Jobin2019} confirm the pattern across 84 AI ethics guidelines: high-level principles converge while implementation diverges. Before proposing a framework, we ask what the instruments already in force actually require of military AI, and of whom.

\subsection{Existing Governance Instruments and Their Limits}
\label{ssec:existing_instruments}

Military AI falls under three regimes. They were written separately, and none of them was written for it. This section makes one claim about them: the rules that bind govern the wrong thing. They govern a weapon, a platform, a system, and they treat the learned model as one more piece of software inside it. The argument has three steps. Two of them are about how far the rules reach. The third is about what the rules take as their object. That third one is the point of this section, because it holds even where a binding rule does apply.

The civil regime is the most developed. The NIST AI Risk Management Framework~\cite{nist2023airmf} organizes risk work around four functions: govern, map, measure, and manage. The EU AI Act~\cite{euaiact2024} goes further and turns a scale of risk into duties that can be enforced against high-risk systems. The military regime is narrower. U.S.\ DoD Directive 3000.09~\cite{dod300009} sets out how autonomous and semi-autonomous weapon systems are reviewed and approved. The NATO AI Strategy~\cite{nato2021aistrategy} commits members to three principles: systems should be governable, traceable, and reliable. The humanitarian regime only states a position. The ICRC~\cite{icrc2021autonomous} asks states to agree on rules for human control over the use of force.

The first limit is about who is covered. The EU AI Act does not apply to systems used only for military, defense, or national security purposes (Art.~2(3))~\cite{euaiact2024}. The exclusion is smaller than it looks. Recital~24 says that a dual-use system, or a military system reused for civilian work, comes back under the Act~\cite{euaiact2024}. Still, the effect is that the one instrument with real enforceable duties does not reach purely military systems, and covers the dual-use middle ground (predictive logistics, ISR, decision support) only in patches. The second limit is about what is covered. DoDD~3000.09 is the one binding rule written for the military, and it applies only to weapon systems~\cite{dod300009}. Logistics, medical triage, intelligence analysis, and cyber operations sit outside it, even when what they produce leads to someone being killed. NIST, NATO, and the ICRC cover the whole range, but as advice or as a stated principle, not as a requirement~\cite{nist2023airmf,nato2021aistrategy,icrc2021autonomous}.

Both of those limits are about coverage, and both could be closed by widening a perimeter. The third one cannot, because it is about what the rules govern rather than how far they reach. To see it, take the case that favors the current regime most: a system that sits squarely inside the strongest binding rule.

Consider an air-defense turret covered by DoDD~3000.09. Its object-detection model reads radar and camera returns and decides which aircraft to fire on. The directive covers the turret, not the model. So the model gets reviewed only as one part of the weapon, and every requirement that follows is attached to the turret. Three consequences follow. The rest of this subsection takes them one at a time.

The first is that the rules track where the model sits rather than what it does. Move the same model into a tool that flags vehicles for a human analyst and it becomes ordinary intelligence software, bound by nothing. It still makes the same mistakes at the same rate. Only the cost of a mistake has changed.

The second is that the checks that do exist ask the wrong question. Checks do happen when the model is updated. New software counts as a new baseline, and that triggers the same re-qualification any other component change would. But the check runs as configuration management, and it tests the model against the specification the platform was approved against. That kind of check assumes two things are enough to tell you how a system will behave in the field: a test campaign of finite size, and a look at how the thing was built. For control logic that follows fixed rules, the assumption holds. It breaks for a model that can get an input wrong even when that input looks no different from the ones it gets right, and whose behavior comes from training data and weights rather than rules an engineer wrote down. The check runs, the system passes, and what you learn is not what the check was built to tell you.

The third is that some failures arrive with no check at all. As the inputs the model sees in the field drift away from the ones it was trained on, the model gets worse while the approved baseline stays exactly the same. Nothing changes on paper, so there is no event for the rules to hang on.

Together these leave a gap in accountability. When the turret fires on the wrong aircraft, the mount, the fire-control logic, the interlocks, and the sensors can each be found within specifications. The configuration can be found compliant. And the error that decided the outcome sits in a classification that no requirement in the regime was written to constrain.

The pattern is not limited to weapons. Existing rules govern a thing that is built, fielded, and approved as a unit, and treat the model as one more piece of software inside it. But a model carries a decision whose stakes are set by where it is installed, and a behavior that can change while the platform stays the same. Governing the platform leaves that decision free. Better rules have to begin by saying what they should require, and of what. The next section builds that list.

% ==========================================
%                BACKGROUND
% ==========================================

% ==========================================
%                 PART3
% ==========================================
\section{Design Principles for Accountable Military AI}
\label{sec:design_principles}

The previous section ended on a missing list: what should good rules require of a military AI? Before writing it, we have to ask where such rules can hold. One answer is to put them inside the model itself.

At the current state of the art, we do not know how to reliably encode ethics into learning systems in a way that remains robust across diverse contexts, adversarial manipulation, and distribution shifts. Arkin~\cite{Arkin2009} proposed the most developed attempt at an ``ethical governor'': a computational architecture that would enforce IHL constraints (distinction, proportionality, prohibition on perfidy) at the system level. The approach demonstrates that certain hard constraints can be implemented as decision-theoretic filters, but it also reveals the limits of formalization. Moral judgment involves interpretation, competing values, and the capacity to recognize when rules should be broken. No loss function captures the principle of proportionality. No training dataset encodes the full meaning of distinction. Attempts to formalize ethics into algorithmic constraints produce brittle systems that satisfy the letter of a rule while violating its spirit, or that fail unpredictably outside their training distribution. Rather than operating under the illusion that morality can be optimized as a mathematical loss function, we propose to \emph{proceduralize} ethics, encoding ethical constraints into institutions, certification regimes, standard operating procedures, and technical safeguards. A meta-analysis of 84 AI ethics guidelines by Jobin \textit{et al.}~\cite{Jobin2019} reveals broad convergence on transparency, fairness, non-maleficence but persistent divergence on implementation. This gap between principle and practice is precisely what proceduralization is designed to close.

This pragmatic stance is not a permanent renunciation of value-aligned design. As alignment research matures, a hybrid approach will become viable, combining limited algorithmic constraints (calibrated uncertainty, forced abstention, hard safety interlocks) with rigorous external procedural controls (mandatory audits, authorization gates, traceability). We develop this approach through six components: actionable accountability structures (\ref{ssec:actionable_accountability}), adversarial evaluation (\ref{ssec:adversarial_evaluation}), a NATO standardization proposal (\ref{ssec:nato_standard}), interface design principles for preserving human judgment (\ref{ssec:assistant_principle}), tiered deployment thresholds (\ref{ssec:deployment_thresholds}) and testing it against eight operational cases (\ref{ssec:operational_precedents}).

\subsection{Actionable Accountability}
\label{ssec:actionable_accountability}

If responsibility is dispersed across many hands, governance must make it actionable. The objective is to ensure that (i)~every high-stakes decision has an identifiable human authorizer, (ii)~every deployed system has an identifiable human owner, and (iii)~every critical failure can be reconstructed and attributed through evidence.

\noindent We propose three complementary requirements.

\runinhead{Named accountability roles.} For each operational AI capability, states should designate a \emph{System Owner} (responsible for lifecycle risk management), an \emph{Operational Authorizer} (responsible for each mission-level activation), an \emph{Operational User} (the operator who interacts with the system in real time, responsible for exercising judgment on its outputs, reporting anomalies, and invoking override or abort procedures), and an \emph{Independent Evaluator} (responsible for certification and periodic reassessment). These roles must be separated to reduce conflicts of interest, and their responsibilities written into doctrine rather than treated as informal practice.

\runinhead{Traceability by default.} Accountability requires investigability. Systems should maintain tamper-evident logs capturing model version and weights hash, data pipeline versioning, sensor provenance (device serial number and certificate identifier), a hash of sensor inputs, uncertainty estimates, operator interactions (overrides, approvals, aborts), and timing information. The design goal is a factual record sufficient for post hoc review and legal scrutiny.

\runinhead{Technical enforceability of oversight.} Whenever a system is used in an operation that can produce harm, the architecture should enforce stage-gated approvals, abort capability, and fallback modes. If substantive human control is a requirement, it must be implemented as a control surface that the operator can exercise under time pressure.

\medskip\noindent
These measures do not eliminate the many-hands problem, but they transform distributed decision-making into distributed obligation, reducing the conditions for scapegoating by making responsibilities explicit, auditable, and enforceable.

\subsection{Adversarial Evaluation}
\label{ssec:adversarial_evaluation}

Accountability structures establish who must answer for a decision, but not whether the underlying system is robust enough to warrant deployment. Evaluation and certification form the second pillar of proceduralized ethics.

The standard approach relies on benchmarks with known ground-truth labels. When evaluation criteria are known in advance, developers optimize for those criteria at the expense of broader competence. Goodhart~\cite{Goodhart1975} first identified this dynamic in monetary policy; Strathern~\cite{Strathern1997} generalized it: ``when a measure becomes a target, it ceases to be a good measure.'' The pattern is well-documented in ML. Dozens of optimizers claim to outperform AdamW on standard benchmarks, yet almost none see adoption, because comparisons rely on asymmetric hyperparameter tuning, favorable evaluation conditions, and unreported negative results~\cite{wen2025fantasticpretrainingoptimizers}. As Jordan argues in his analysis of the Muon optimizer~\cite{jordan2024muon}, the only credible evidence of superiority is success under truly competitive conditions. Recht \textit{et al.}~\cite{Recht2019} provide a striking illustration at the dataset level: after replicating the ImageNet test set creation process from scratch, they observed accuracy drops of 11-14\% across a wide range of classifiers, with no change in the underlying data distribution. Performance on a fixed benchmark, even a well-curated one, does not reliably predict generalization.

For military AI, the same dynamic applies with higher stakes. If defense contractors are evaluated on disclosed benchmarks, optimization will target those benchmarks, producing systems that perform well in controlled demonstrations but may fail in operational environments. A natural countermeasure is adversarial auditing with undisclosed evaluation criteria. Benchmarks used to certify military AI should be kept secret, accessible only to authorized evaluators, and rotated regularly. The existence of certain evaluation sets should itself be classified. The evaluation process should remain transparent and standardized; what must stay hidden is the evaluation content. Transparency about process combined with secrecy about content channels the incentive structure toward competence rather than benchmark-specific optimization.

Evaluation should also include adversarial robustness testing. As discussed in \ref{ssec:opacity}, adversarial patches~\cite{brown2018adversarialpatch} and targeted perturbations~\cite{8578273} can cause high-confidence misclassification on inputs that appear benign to human observers. Findings ~\cite{Carlini2017, Madry2018} suggest that robustness under adversarial conditions is a more informative criterion than accuracy on clean data. Hendrycks and Dietterich~\cite{Hendrycks2019} demonstrated this concretely with ImageNet-C, showing that top-performing classifiers degrade severely under common corruptions (noise, blur, weather, digital artifacts) despite high clean-data accuracy. For military systems operating in contested and unpredictable environments, evaluation must reflect these conditions.

A complementary dimension is negative testing. Standard evaluation emphasizes positive cases (does the model correctly identify targets it should identify?), but the converse matters equally: does the model correctly reject inputs it should reject? A classifier trained on military vehicles may learn to associate camouflage patterns with the positive class, triggering false positives on civilian trucks while missing military vehicles in unexpected color. For military AI, negative testing directly operationalizes the principle of distinction. Evaluation should include civilian objects sharing features with military targets, military targets with atypical appearances, scenarios involving surrender or medical evacuation, and inputs at the boundary of the training distribution.

\subsection{Toward a NATO Evaluation Standard}
\label{ssec:nato_standard}

NATO already possesses the normative infrastructure to standardize how sensor data is captured, formatted, and exchanged. Its 2021 AI Strategy~\cite{nato2021aistrategy} commits member states to responsible AI development based on principles of governability, traceability, and reliability, but stops short of specifying certification mechanisms. What NATO lacks is an equivalent framework for certifying the algorithms that consume this data. Table~\ref{tab:stanag_landscape} summarizes the current landscape.

\begin{table}
    \caption{NATO standardization landscape for AI-relevant capabilities. Sensor-layer standards are mature; algorithmic and autonomy layers remain under development. Study-stage STANAGs have not been ratified; objectives are drawn from the NATO Standardization Office DCRA Report~\cite{nato_dcra}.}
    \label{tab:stanag_landscape}
    \centering
    \small
    \setlength{\tabcolsep}{4pt}
    \renewcommand{\arraystretch}{1.3}
    \begin{tabularx}{\textwidth}{l l X l}
        \toprule
        \textbf{Layer} & \textbf{STANAG} & \textbf{Purpose} & \textbf{Status} \\
        \midrule
        Sensor & 4607 & Ground Moving Target Indicator (GMTI) data transmission~\cite{stanag4607} & Promulgated \\
        Sensor & 4609 & NATO Digital Motion Imagery (Full Motion Video (FMV) with Video Moving Target Indicator (VMTI) metadata) & Promulgated \\
        Sensor & 4676 & Intelligence, Surveillance and Reconnaissance (ISR) tracking data and trajectory exchange & Promulgated \\
        Sensor & 4579 & Battlefield Target Identification (electronic IFF) & Promulgated \\
        Sensor & 5527 & Friendly Force Tracking interoperability & Promulgated \\
        Algorithmic & 5653 & Core Data Framework (common metadata schemas)~\cite{stanag5653} & Study stage \\
        Algorithmic & 5670 & Federated Data Catalogue (decentralized metadata discovery)~\cite{stanag5670} & Study stage \\
        Algorithmic & 5669 & Neural network and deep learning model exchange~\cite{stanag5669} & Study stage \\
        Autonomy & 4817 & Multi-domain unmanned platform command and control & Study stage \\
        \bottomrule
    \end{tabularx}
\end{table}

STANAG~5669 is the most ambitious of the study-stage efforts: it targets the exchange of trained neural network models between nations independently of training-time software and hardware~\cite{stanag5669}. In principle, this would enable inference sharing, cross-domain fine-tuning, and multilateral model development. In practice, however, the exchange of trained military AI models faces a fundamental obstacle. Military AI systems are developed on top of national doctrine, encoding tactical assumptions, operational priorities, and decision heuristics specific to each nation's armed forces. Sharing a trained model amounts to exposing that doctrine. An adversary with access to the model can reverse-engineer the decision logic it embodies and develop targeted countermeasures, for instance by training a system specifically designed to exploit the doctrinal patterns encoded in the weights. This concern, raised in discussions with Lieutenant-Colonel J\'er\^ome Ranc at the Human Factors Air Operations Laboratory (Centre d'Expertise A\'erienne Militaire / Air Warfare Center), suggests that unrestricted model exchange between allies will remain impractical for the foreseeable future.

What is both feasible and needed is a standardized evaluation protocol. We propose that NATO member states develop such a protocol, formalized as a STANAG. The protocol should be structured in two parts. The first, intended for states and evaluation authorities, specifies:
\begin{itemize}
    \item Rules governing benchmark secrecy, including classification levels, access controls, and rotation schedules.
    \item Evaluation methodology: which metrics are measured, how thresholds are determined, and pass/fail criteria.
    \item Prerequisites for deployment authorization, including documentation, audit trails, and fallback mechanisms.
    \item Governance structures for independent auditing bodies.
\end{itemize}

A critical design question concerns the transparency of evaluation criteria. If developers know exactly which metrics are used, they optimize for those metrics (Goodhart's Law again). If metrics are entirely opaque, developers cannot adequately prepare. A middle path is available: the categories of metrics (robustness, calibration, latency) can be made public while the specific implementations, thresholds, and weighting schemes remain classified. This does not eliminate the Goodhart risk, but it channels optimization toward relevant properties rather than the idiosyncrasies of a known test suite.

The second part, intended for developers, specifies:
\begin{itemize}
    \item The category of the model and its intended operational use.
    \item The training dataset, or a statistical characterization sufficient to compute distributional properties.
    \item Compliance with a unified annotation standard per modality and task.
\end{itemize}

Full dataset disclosure would enable auditors to detect distributional biases and spurious correlations, but training data often constitutes proprietary assets. A workable compromise requires developers to provide statistical summaries (class distributions, domain coverage, annotation quality metrics) sufficient for effective audit, complemented by periodic sample-level inspections under confidentiality agreements. Summary statistics have limited power to reveal certain bias classes; sample-level audits provide the necessary complement.

Interoperability demands annotation standardization. The proliferation of labeling formats (Pascal VOC, COCO, custom schemas) creates friction and enables format-dependent inconsistencies. A STANAG for military AI should mandate a canonical annotation format per modality and task, with open-source conversion tools for legacy formats. Table~\ref{tab:annotation_modalities} illustrates the breadth of modalities involved and their varying degrees of standardization maturity.

\begin{table}
    \caption{Modalities, representative tasks, structural annotations, and modality-specific metadata relevant to military AI evaluation. The standardization column reflects the availability of public benchmarks and shared annotation schemas. Cross-cutting metadata requirements (provenance, uncertainty, operational context, versioning) apply to all modalities and are discussed in the text.}
    \label{tab:annotation_modalities}
    \centering
    \small
    \setlength{\tabcolsep}{4pt}
    \renewcommand{\arraystretch}{1.3}
    \begin{tabularx}{\textwidth}{l X X X c}
        \toprule
        \textbf{Modality} & \textbf{Representative Tasks} & \textbf{Structural Annotations} & \textbf{Modality-Specific Metadata} & \textbf{Std.} \\
        \midrule
        Imagery (EO/IR) & Detection, classification, segmentation, tracking & Bounding boxes, pixel masks, class labels, object IDs, occlusion flags & Sensor type, Ground Sample Distance, altitude, weather, illumination & $\bullet\bullet\bullet$ \\
        Video / FMV & Activity recognition, target tracking, event detection & Temporal segments, action labels, trajectory annotations & Frame rate, compression, camera motion, stabilization & $\bullet\bullet\circ$ \\
        Radar / SAR & Automatic target recognition, change detection, ship detection & Target chips, detection masks, class labels & Polarization, frequency band, incidence angle, range resolution & $\bullet\circ\circ$ \\
        Geospatial & Change detection, terrain classification, infrastructure mapping & Geo-referenced polygons, temporal stamps, attribution labels & Coordinate Reference System (CRS), spatial resolution, temporal coverage, collection geometry & $\bullet\bullet\circ$ \\
        3D & Object detection, scene reconstruction, ground segmentation & 3D bounding boxes, point-wise class labels, mesh annotations & Sensor origin (LiDAR, stereo, Structure from Motion (SfM)), point density, registration accuracy & $\bullet\bullet\circ$ \\
        Signals (SIGINT) & Emitter classification, signal detection, protocol identification & Time-frequency annotations, emitter IDs, protocol labels & Center frequency, bandwidth, Signal-to-Noise Ratio (SNR), collection geometry & $\bullet\circ\circ$ \\
        Text & Entity extraction, relation extraction, event detection & Span annotations, entity types, relation labels & Source type, language, reliability rating, collection date & $\bullet\bullet\bullet$ \\
        Audio & Speaker ID, speech recognition, acoustic event detection & Transcripts, speaker labels, temporal boundaries & Sample rate, SNR, channel count, recording environment & $\bullet\bullet\circ$ \\
        Multi-modal & Cross-modal fusion, joint entity resolution & Cross-modal alignment, joint entity references, per-modality annotation layers & Temporal synchronization, spatial co-registration, modality weighting & $\bullet\circ\circ$ \\
        \bottomrule
    \end{tabularx}
    \footnotetext{Standardization maturity (Std.): $\bullet\bullet\bullet$ = established benchmarks and public annotation schemas; $\bullet\bullet\circ$ = partial standardization or limited military-specific coverage; $\bullet\circ\circ$ = predominantly ad hoc or classified formats.}
\end{table}

Beyond the modality-specific metadata catalogued in Table~\ref{tab:annotation_modalities}, a military annotation standard requires cross-cutting provisions: data provenance and chain of custody, uncertainty encoding (annotator confidence, inter-annotator agreement, ambiguity flags), operational context (scenario type, rules-of-engagement applicability), and versioning (schema identifiers, revision histories). The standard should include a governance mechanism for schema evolution so that the format remains current without sacrificing backward compatibility.

\subsection{The Assistant Principle: Preserving Human Judgment}
\label{ssec:assistant_principle}

If AI systems are to support rather than supplant human decision-making, the interface between operator and algorithm becomes a critical design surface. Norman's concept of affordance~\cite{norman1999affordance} captures the core requirement: the system's perceptible properties should make its capabilities, limitations, and confidence immediately visible, so that the operator's mental model tracks the system's actual state.

The central risk is miscalibrated trust. The interface should make uncertainty perceptually salient so that trust is continuously recalibrated by the display itself. Risko and Gilbert~\cite{risko2016cognitive} frame this as selective cognitive offloading: the system absorbs computational burden (sensor fusion, pattern detection) while freeing attentional resources for judgment, without introducing competing demands.

Where operational constraints permit, inherently transparent model classes (generalized additive models~\cite{lou2012}, decision trees) should be preferred for high-stakes decisions. For deep models, post hoc explanations (saliency maps, SHAP values) provide diagnostic signals but do not guarantee faithful access to internal reasoning~\cite{Lipton2018-dy}. In our framework, post hoc explanations are treated as audit artifacts: they support traceability and review but do not by themselves confer legitimacy on a decision.

These principles suggest a design philosophy we call the \emph{hybrid trajectory}: rather than embedding ethics directly in the model, the goal is to combine procedural safeguards with internal technical guardrails and to expand machine autonomy incrementally as reliability is demonstrated under operational conditions. The trajectory rests on three mechanisms:

\begin{enumerate}
  \item \textbf{Calibrated uncertainty and abstention.} The system estimates and communicates its confidence. Below a task-specific threshold, it abstains and defers to the operator.

  \item \textbf{Hard interlocks.} Actions outside the authorized operational envelope (ROE\footnote{Rules of Engagement} violations, geographic exclusion zones) are made mechanically impossible, independently of software-level controls.

  \item \textbf{Progressive autonomy.} The boundary between autonomous operation and human deferral is set conservatively at initial deployment and widened only when traceability records (\ref{ssec:actionable_accountability}) demonstrate sustained reliability under operational conditions. Crucially, this expansion of autonomy always remains bounded by the hard interlocks defined in Point 2.
\end{enumerate}

The hybrid trajectory is not a fixed architecture but a temporal process: each expansion of machine autonomy is contingent on demonstrated reliability, and procedural safeguards define the permissible boundary at every stage. This logic directly motivates the tiered deployment framework developed next.

\subsection{Tiered Deployment Thresholds}
\label{ssec:deployment_thresholds}

Not all military AI systems warrant the same oversight. A tiered classification organized by increasing operational risk and decreasing reversibility determines the minimum procedural safeguards at each stage of development, certification, and fielding. This logic is consistent with the risk-based architecture of the EU AI Act~\cite{euaiact2024}, the OECD Recommendation on Artificial Intelligence~\cite{oecd2019ai}, and the NIST AI Risk Management Framework~\cite{nist2023airmf}, the differentiated treatment of autonomous weapon systems in U.S.\ DoD Directive 3000.09~\cite{dod300009}, and the ICRC position that the acceptability of autonomy should be assessed relative to the nature and context of the task~\cite{icrc2021autonomous}. Where the EU AI Act classifies by application domain, our tiering classifies by operational consequence and reversibility, reflecting the distinct risk structure of military operations.

We formalize this classification along five measurable dimensions (Table~\ref{tab:tier_formalization}):
\begin{itemize}
    \item \textbf{Lethality potential:} whether the system's outputs can cause death directly, indirectly through downstream decisions, or not at all.
    \item \textbf{Reversibility:} whether a decision can be recalled or corrected after execution.
    \item \textbf{Autonomy level:} the degree of human involvement in the decision loop.
    \item \textbf{Speed of effect:} the temporal window available for human intervention.
    \item \textbf{Scope of impact:} the organizational and strategic breadth of consequences.
\end{itemize}
Tier assignment is determined by the combination of these dimensions; escalation along any single dimension triggers reassessment under the re-certification procedure described below.

\begin{table}
    \caption{Formalized tier classification of military AI capabilities. Each tier is characterized by a profile across five measurable dimensions. Escalation on any single dimension may trigger reclassification.}
    \label{tab:tier_formalization}
    \centering
    \small
    \setlength{\tabcolsep}{4pt}
    \renewcommand{\arraystretch}{1.3}
    % \begin{tabularx}{\textwidth}{l X X X X X}
    \begin{tabularx}{\textwidth}{@{}R{1.30} R{0.75} R{1.05} R{1.10} R{1.00} R{0.80}@{}}
        \toprule
        \textbf{Tier (capability)} &
        \textbf{Lethality} &
        \textbf{Reversibility} &
        \textbf{Autonomy} &
        \textbf{Speed of effect} &
        \textbf{Scope} \\
        \midrule
        1: Predictive logistics & None & Full & Advisory & Human-time & Individual \\
        2: Medical / maintenance & None & Partial & Advisory & Human-time & Individual \\
        3: ISR / surveillance & Indirect & Partial & Semi-autonomous & Accelerated & Unit \\
        4: Target nomination & Indirect & Partial & Semi-autonomous & Accelerated & Strategic \\
        5: Cyber operations & Indirect & Irreversible & Autonomous & Machine-speed & Strategic \\
        6: Lethal engagement & Direct & Irreversible & Autonomous & Machine-speed & Strategic \\
        \bottomrule
    \end{tabularx}
    \footnotetext{\textbf{Lethality:} None = no causal path to harm; Indirect = outputs feed decisions that may cause harm; Direct = system can apply force.}
    \footnotetext{\textbf{Reversibility:} Full = decision can be recalled without residual effect; Partial = correction possible but downstream consequences may persist; Irreversible = effects cannot be undone.}
    \footnotetext{\textbf{Autonomy:} Advisory = human decides; Semi-autonomous = system acts, human approves or vetoes; Autonomous = system acts without per-action human approval.}
    \footnotetext{\textbf{Speed:} Human-time = hours to days; Accelerated = seconds to minutes; Machine-speed = milliseconds.}
    \footnotetext{\textbf{Scope:} Individual = single system or process; Unit = operational-unit-level consequences; Strategic = cross-domain or national-level consequences.}
\end{table}

At minimum, each tier should specify:
\begin{itemize}
    \item \textbf{Pre-deployment assurance:} depth of independent testing (including adversarial and negative testing) and re-certification frequency.
    \item \textbf{Authorization gates:} who approves activation, at what command level, and whether stage-gated approvals are required.
    \item \textbf{Fallback and containment:} required fail-safe modes (safe stop, degraded operation, manual override) and technically enforced function blocks.
    \item \textbf{Monitoring and traceability:} minimum logging requirements, incident-reporting triggers, and conditions for automatic suspension.
\end{itemize}

Lower tiers (1-2: logistics, medical support) require standard software-engineering assurance and domain-expert validation. At intermediate tiers (3-4: reconnaissance, command support), errors can directly affect situational awareness and may become irreversible once acted upon. These tiers call for independent verification and validation including structured red-teaming, command-level deployment authorization, exposure of confidence estimates to operators in accordance with \ref{ssec:assistant_principle}, and technically guaranteed fallback to manual operation.

Tier~5 covers autonomous cyber operations, whose effects propagate at machine speed~\cite{Horowitz2019}, cross jurisdictional boundaries, and trigger cascading consequences difficult to anticipate. National-authority-level approval should be required. Technical containment must include hard scope limits (target lists, network boundaries, time windows) enforced at the code level. Continuous monitoring with automatic suspension triggers is essential given the latency between misfire and detection in cyberspace.

Tier~6 concerns lethal autonomous weapon systems and raises a more fundamental question: whether a given capability should be deployed at all. The ICRC has called for new internationally agreed rules ensuring human control over the use of force, and has argued that autonomous weapons incapable of IHL compliance should be expressly prohibited~\cite{icrc2021autonomous}. Within the present framework, Tier~6 does not function as authorization to deploy under sufficiently robust safeguards; it marks a threshold of political and legal deliberation.

Even where all safeguards are satisfied, certain configurations may remain impermissible. A system that cannot reliably distinguish combatants from civilians, or that is intended for environments where such distinction is structurally unachievable, should not be deployed regardless of procedural compliance. Procedural governance defines necessary conditions; it does not establish sufficiency.

A system's tier is not fixed at deployment. Operational experience, adversarial adaptation, and incremental upgrades can shift the risk profile. A reconnaissance AI at Tier~3 that acquires target-nomination features drifts toward Tier~4 or beyond; if this migration is not formally recognized, safeguards remain calibrated to a lower risk level. The framework must incorporate periodic re-assessment with mandatory re-certification whenever functional scope or operational context changes materially~\cite{nist2023airmf, dod300009}.

\subsection{From Dimension Profiles to Tiers: A Criticality Score}
\label{ssec:criticality}

Table~\ref{tab:tier_formalization} names six reference profiles, one per tier. A fielded system rarely matches one of them exactly: with three levels on each of five dimensions there are $3^5 = 243$ possible profiles, and the table names six. Tier assignment therefore needs a rule for the other $237$.

We write $x = (x_1, \dots, x_5)$ for a profile, where each $x_d \in \{0, 1, 2\}$ ranks the level of one dimension from least to most severe (for lethality, None~$=0$, Indirect~$=1$, Direct~$=2$, and likewise for the others in the order of Table~\ref{tab:tier_formalization}). We write $A_1, \dots, A_6$ for the six reference profiles.

\runinhead{Why not a weighted sum.} The natural first idea is to weight each dimension and add: $S(x) = \sum_d w_d\, x_d$, then cut the score into six intervals. Two elementary facts rule this out for Table~\ref{tab:tier_formalization}.

\runinhead{Proposition 1 (a weighted sum has no interactions).} Under $S$, raising one dimension by one level changes the score by $w_d$, whatever the other four dimensions are. The cost of autonomy is the same for a system acting in milliseconds as for one acting over days; the cost of scope is the same whether or not the system is lethal. This follows immediately from the form of $S$: the four terms that do not change cancel. In Table~\ref{tab:tier_formalization} the step from Tier~5 to Tier~6 changes lethality alone, so $w_{\text{lethality}}$ is pinned to that single gap, in every context.

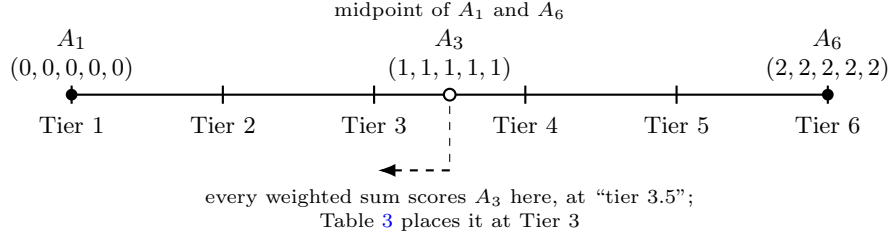
\begin{figure}[t]
    \centering
    \begin{tikzpicture}[font=\small, >=Latex]
        \draw[thick] (0,0) -- (10,0);
        \foreach \p/\t in {0/1, 2/2, 4/3, 6/4, 8/5, 10/6} {
            \draw[thick] (\p,0.12) -- (\p,-0.12) node[below=1pt] {Tier~\t};
        }
        \fill (0,0) circle (2.2pt);
        \node[above=4pt, align=center, inner sep=1pt] at (0,0) {$A_1$\\$(0,0,0,0,0)$};
        \fill (10,0) circle (2.2pt);
        \node[above=4pt, align=center, inner sep=1pt] at (10,0) {$A_6$\\$(2,2,2,2,2)$};
        \draw[thick, fill=white] (5,0) circle (2.2pt);
        \node[above=4pt, align=center, inner sep=1pt] (a3) at (5,0) {$A_3$\\$(1,1,1,1,1)$};
        \node[above=1pt, font=\fontsize{7.5}{9}\selectfont, inner sep=1pt] at (a3.north) {midpoint of $A_1$ and $A_6$};
        \draw[dashed] (5,-0.15) -- (5,-1.0);
        \draw[->, thick, dashed] (5,-1.0) -- (4.05,-1.0);
        \node[below=2pt, align=center, font=\fontsize{7.5}{9}\selectfont] at (5,-1.05) {every weighted sum scores $A_3$ here, at ``tier~3.5'';\\ Table~\ref{tab:tier_formalization} places it at Tier~3};
    \end{tikzpicture}
    \caption{Why a weighted sum of the five ranks cannot reproduce Table~\ref{tab:tier_formalization} with evenly spaced tiers. The Tier~3 profile is the arithmetic midpoint of the Tier~1 and Tier~6 profiles, so any weighted sum places it halfway between them, between Tiers~3 and~4.}
    \label{fig:midpoint}
\end{figure}

\runinhead{Proposition 2 (a weighted sum cannot space the tiers evenly).} The Tier~3 profile $A_3 = (1,1,1,1,1)$ is the exact midpoint of the Tier~1 profile $A_1 = (0,0,0,0,0)$ and the Tier~6 profile $A_6 = (2,2,2,2,2)$. Any weighted sum therefore scores it exactly halfway between Tiers~1 and~6, that is, at ``tier~3.5'' on an evenly spaced scale (Fig.~\ref{fig:midpoint}), whereas the table places it at Tier~3. Formally, $S(A_3) = \tfrac12\bigl(S(A_1) + S(A_6)\bigr)$ for every choice of weights. Consequently no weights make the six tiers equally spaced. Worse, the gap from Tier~2 to~3 equals the gap from Tier~3 to~4 plus the gap from Tier~5 to~6 plus $w_{\text{autonomy}} + w_{\text{speed}}$, so the largest gap between consecutive tiers is at least twice the smallest, with equality only when autonomy and speed carry zero weight.

Both facts concern the same restriction: a weighted sum of ranks treats the two steps of a dimension as equally costly and the five dimensions as independent. Neither is credible here. Going from Indirect to Direct lethality is not the same step as going from None to Indirect; and speed of effect matters because it shortens the window in which a human can intervene, which is only relevant if the system acts on its own.

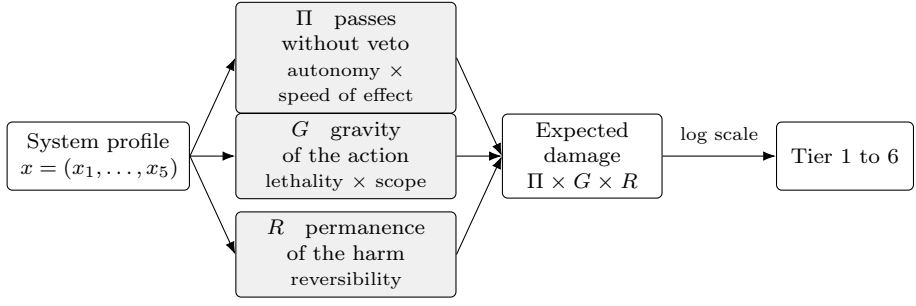
\begin{figure}[t]
    \centering
    \begin{tikzpicture}[font=\fontsize{8}{9.6}\selectfont, >=Latex,
        box/.style={draw, rounded corners=2pt, align=center, minimum height=9mm, inner sep=3pt},
        fac/.style={box, fill=gray!12, text width=27mm}]
        \node[box, text width=22mm] (prof) {System profile\\ \mbox{$x = (x_1,\dots,x_5)$}};
        \node[fac, right=6mm of prof, yshift=13mm] (pi) {$\Pi$ \; passes without veto\\ \fontsize{7}{8.4}\selectfont autonomy $\times$ speed of effect};
        \node[fac, right=6mm of prof] (g) {$G$ \; gravity of the action\\ \fontsize{7}{8.4}\selectfont lethality $\times$ scope};
        \node[fac, right=6mm of prof, yshift=-13mm] (r) {$R$ \; permanence of the harm\\ \fontsize{7}{8.4}\selectfont reversibility};
        \node[box, right=6mm of g, text width=19mm] (crit) {Expected damage\\ $\Pi \times G \times R$};
        \node[box, right=15mm of crit, text width=16mm] (tier) {Tier $1$ to $6$};
        \draw[->] (prof.east) -- (pi.west);
        \draw[->] (prof.east) -- (g.west);
        \draw[->] (prof.east) -- (r.west);
        \draw[->] (pi.east) -- (crit.west);
        \draw[->] (g.east) -- (crit.west);
        \draw[->] (r.east) -- (crit.west);
        \draw[->] (crit) -- node[above=1pt, font=\fontsize{7}{8.4}\selectfont] {log scale} (tier);
    \end{tikzpicture}
    \caption{Structure of the criticality score. The five dimensions of Table~\ref{tab:tier_formalization} enter through three factors that multiply; the tier is read on a logarithmic scale, so that moving up one tier always means the same multiplicative increase in expected damage. Dimensions within a factor may reinforce each other; dimensions in different factors do not.}
    \label{fig:criticality_structure}
\end{figure}

\runinhead{The score we adopt.} We read a tier as a level of \emph{expected damage}: how likely a wrong action is to take effect, times how bad it is, times how long it lasts. The five dimensions fall into three groups accordingly (Fig.~\ref{fig:criticality_structure}):
\begin{equation}
    \operatorname{crit}(x) = \underbrace{\Pi(\text{autonomy}, \text{speed})}_{\text{passes without veto}}
    \times \underbrace{G(\text{lethality}, \text{scope})}_{\text{gravity of the action}}
    \times \underbrace{R(\text{reversibility})}_{\text{permanence of the harm}} .
    \label{eq:criticality}
\end{equation}
$\Pi$ captures the \emph{veto window}: a system that only advises leaves the decision, and hence the veto, to a human, whatever its speed; a system that acts on its own can still be stopped if its effects unfold over hours, but not if they unfold in milliseconds. $G$ measures the gravity of the action itself, in which scope matters more when the action is lethal. $R$ records whether the harm can be undone once it has occurred; reversibility is a property of the damage, not of the probability that a wrong action is taken.

This form has two consequences. First, within each factor the two dimensions can reinforce each other, which Proposition~1 said a weighted sum cannot do. Second, across factors the score multiplies, so that on the tier scale the three contributions add: the cost of a lethality step is the same whatever the autonomy level.

\runinhead{What the factors are, and how a product becomes a sum.} $\Pi$, $G$ and $R$ are small tables of declared values (Table~\ref{tab:criticality_tables}). Moving up one tier always means multiplying expected damage by the same constant factor, call it $\beta$. A level that multiplies expected damage by $\beta$ is therefore worth one tier, a level that multiplies it by $\beta^{2}$ is worth two, and in general a factor $F$ is worth $\log_\beta F$ tiers. We record every table entry in these units, so that $g = \log_\beta G$, $\pi = \log_\beta \Pi$ and $r = \log_\beta R$. Counted in tiers, factors that multiply become numbers that add, as decibels turn ratios of power into sums. Taking logarithms of Eq.~\eqref{eq:criticality} gives
\begin{align}
    \text{score}(x) \;&=\; \log_\beta \operatorname{crit}(x) \notag \\
                    &=\; g[\text{lethality}, \text{scope}] + \pi[\text{autonomy}, \text{speed}] + r[\text{reversibility}],
    \label{eq:score}
\end{align}
and the tier is one plus the nearest integer to the score. The numerical value of $\beta$ never needs to be fixed: only tier units enter the tables and the rule. The six reference profiles of Table~\ref{tab:tier_formalization} score $0.00$, $0.65$, $2.32$, $2.68$, $3.99$ and $5.34$, and therefore fall in Tiers~1 to~6 as required. Raising any single dimension never lowers the score, so escalation on one dimension can only maintain or raise the tier, as the re-certification rule above assumes.

\begin{table}[t]
    \caption{The three factors of the criticality score, expressed in tiers: an entry of $1.0$ means that this level multiplies expected damage by the factor $\beta$ separating two consecutive tiers, an entry of $2.0$ by $\beta^{2}$, and so on. The tier of a profile is one plus the nearest integer to the sum of its three entries. Profiles combining Direct lethality with Full reversibility are treated as inadmissible: a system that can apply force cannot have fully recallable effects.}
    \label{tab:criticality_tables}
    \centering
    \small
    \setlength{\tabcolsep}{4pt}
    \renewcommand{\arraystretch}{1.3}
    \begin{tabularx}{\textwidth}{@{}R{1.3} Z{0.9} Z{0.9} Z{0.9}@{}}
        \toprule
        \multicolumn{4}{@{}l}{$g$: gravity of the action} \\
        \textbf{Lethality} $\backslash$ \textbf{Scope} & Individual & Unit & Strategic \\
        \midrule
        None     & 0.00 & 0.10 & 0.39 \\
        Indirect & 0.75 & 1.03 & 1.39 \\
        Direct   & 2.10 & 2.38 & 2.74 \\
        \midrule
        \multicolumn{4}{@{}l}{$\pi$: the action passes without veto} \\
        \textbf{Autonomy} $\backslash$ \textbf{Speed} & Human-time & Accelerated & Machine-speed \\
        \midrule
        Advisory        & 0.00 & 0.00 & 0.00 \\
        Semi-autonomous & 0.30 & 0.64 & 0.96 \\
        Autonomous      & 0.60 & 0.95 & 1.30 \\
        \midrule
        \multicolumn{4}{@{}l}{$r$: permanence of the harm} \\
        \textbf{Reversibility} & Full & Partial & Irreversible \\
        \midrule
                        & 0.00 & 0.65 & 1.30 \\
        \bottomrule
    \end{tabularx}
\end{table}

\runinhead{How the values were set.} The entries of Table~\ref{tab:criticality_tables} were fitted to no data: they satisfy six constraints, each of which can be accepted or rejected on its own by an expert panel, and within what those constraints leave open they were chosen so that no admissible profile sits close to a tier boundary (the closest is a tenth of a tier away):
\begin{enumerate}
    \item \textbf{Veto window.} The Advisory row of $\pi$ is zero: if a human decides, the speed of effect is immaterial. Below that row, speed counts for more as autonomy grows, and conversely.
    \item \textbf{Gravity.} In $g$, scope counts for more as lethality grows, and lethality for more as scope grows.
    \item \textbf{Admissibility.} A system with Direct lethality cannot have Full reversibility; the $27$ such profiles are excluded, leaving $216$.
    \item \textbf{Fidelity to Table~\ref{tab:tier_formalization}.} Each reference profile falls inside its own tier band, at a safe distance from the edges, without being forced to the centre of the band. Forcing the centre would assert that the joint rise of four dimensions (Tier~2 to~3) is worth the rise of scope alone (Tier~3 to~4), a precision the table does not carry.
    \item \textbf{Lethality as a threshold.} Moving from None to Direct lethality adds at least two tiers, at every scope.
    \item \textbf{No silent dimension.} Every step on every dimension costs at least a tenth of a tier, the two steps of a dimension are within a factor of three of each other, and an autonomous system acting at machine speed sits at least $1.2$ tiers above an advisory one.
\end{enumerate}
A panel should debate these six statements, not the sixteen free entries of Table~\ref{tab:criticality_tables}. On one point Table~\ref{tab:tier_formalization} leaves them little room. It grants a single tier to the joint rise of reversibility, autonomy and speed to their maximum (Tier~4 to~5), but also a single tier to the rise of scope alone (Tier~3 to~4). Any score faithful to the table therefore weighs autonomy lightly relative to scope, and constraint~6 keeps irreversibility from disappearing in the trade-off. Accepting this, or revising the Tier~4 and Tier~5 reference profiles, is the panel's decision.

\subsection{Operational Precedents: Counterfactual Analysis}
\label{ssec:operational_precedents}

The governance architecture proposed above is intended to be more than a normative exercise. To assess whether its mechanisms would make a practical difference, we apply the framework to eight documented cases spanning Tiers~3-6 and the period 1988-2025 (Table~\ref{tab:case_summary}). For each governance component, we identify cases where its absence contributed to documented harm and examine what the framework would have required. Not all cases can be fully verified from public information; the analysis should be read as systematic counterfactual assessment. Counterfactual reasoning demonstrates that the framework's mechanisms map onto documented failure classes, but it cannot establish that their presence would have prevented harm with certainty. Prospective evaluation through integration into certification procedures remains the necessary next step. These limitations acknowledged, the analysis serves a specific purpose: to show that each governance component addresses a failure mode that has occurred in practice, and that their combination defines a governance posture substantially more demanding than any currently in force.

The evidential basis varies across cases. Some are documented through official investigations~\cite{fogarty1988vincennes, dsb2005patriot, DoDIG2022Maven}. Others rest on UN Panel of Experts reports~\cite{unpanel2021libya}, defense analyses~\cite{cepa2024irondome}, or independent technical forensics~\cite{langner2011stuxnet}. The Lavender and Gospel cases rely primarily on investigative journalism~\cite{abraham2024lavender, hrw2024lavender, hrw2024gaza} and have not been subject to official inquiry. We draw on them as the best available evidence while acknowledging that future disclosures may revise the factual record. Table~\ref{tab:case_summary} summarizes each case. Table~\ref{tab:case_matrix} maps which framework dimensions were satisfied, absent, or partially met.

\begin{table}
    \caption{Summary of eight operational cases used for counterfactual framework analysis, ordered chronologically. Tier assignments follow the classification in Table~\ref{tab:tier_formalization}.}
    \label{tab:case_summary}
    \centering
    \small
    \setlength{\tabcolsep}{4pt}
    \renewcommand{\arraystretch}{1.2}
    % \begin{tabularx}{\textwidth}{l l c l l X}
    \begin{tabularx}{\textwidth}{@{}R{0.50} l c R{0.70} R{0.81} R{1.99}@{}}
        \toprule
        \textbf{Case} &
        \textbf{Year} &
        \textbf{Tier} &
        \textbf{System type} &
        \textbf{Primary deficiency} &
        \textbf{Documented consequence} \\
        \midrule
        USS Vincennes & 1988 & 6 & Aegis-assisted engagement & Interface design, trust miscalibration & 290 civilians killed (Iran Air 655) \\
        Patriot friendly fire & 2003 & 6 & Autonomous air defense & Automation bias, IFF failure & 3 allied aircrew killed \\
        Stuxnet & 2007-10 & 5 & Autonomous cyber weapon & No scope limits, no recall & Spread to ${\sim}$115,000 systems worldwide \\
        Iron Dome & 2011- & 6 & Defensive interception & (Positive case) & ${>}$90\% interception rate \\
        Project Maven & 2017- & 3${\to}$4 & ISR, target nomination & Contractor-state knowledge asymmetry & Oversight deficiencies (DoD IG) \\
        Kargu-2 & 2020 & 6 & Loitering munition & Absence of all safeguards & Autonomous engagement without oversight \\
        Gospel & 2021- & 4 & Structural targeting & Volume vs.\ review quality & 12,000+ targets, review capacity overwhelmed \\
        Lavender & 2023 & 5 & Individual targeting & Performative endorsement & ${\sim}$3,700 est.\ misidentifications \\
        \bottomrule
    \end{tabularx}
\end{table}

\subsubsection{Tiered deployment thresholds.} The tiering framework requires that capabilities be classified by their risk profile and that migration across tiers trigger re-certification. Project Maven illustrates the consequences of unrecognized tier drift. Originally an ISR analysis tool (Tier~3), Maven evolved under Palantir into the Maven Smart System, which includes an AI Asset Tasking Recommender that proposes bomber and munition assignments to targets~\cite{defensescoop2024maven}. By 2025, the contract ceiling exceeded \$1.3~billion and the system had been adopted by NATO Allied Command Operations~\cite{nato2025maven}. This migration from Tier~3 to Tier~4 occurred without public evidence of formal re-assessment. The Kargu-2 represents the inverse failure: a Tier~6 capability (autonomous lethal engagement) deployed without any of the governance requirements that tier demands~\cite{unpanel2021libya}. No political-level authorization, no IHL legal review, no geographic or target-class constraints enforced at the system level. Iron Dome provides the positive counterexample. Although it possesses Tier~6 capability (autonomous launch of interceptor missiles), the system operates within tightly constrained parameters: the target set is incoming projectiles, the decision criterion is ballistic impact-point prediction, and defended zones are politically authorized~\cite{cepa2024irondome}. Iron Dome demonstrates that Tier~6 governance requirements are compatible with effective autonomous operation when the operational envelope is well-defined.

\subsubsection{Actionable accountability.} The framework requires named accountability roles (System Owner, Operational Authorizer, Independent Evaluator) with separated responsibilities. Project Maven's first phase exposed what happens when these roles are absent. Google engineers who raised ethical concerns lacked formal channels to influence deployment decisions; DoD officials who authorized deployment lacked direct access to the underlying code or training data~\cite{DoD2017Maven, DoDIG2022Maven}. Knowledge resided in the contractor, authority in the state, and no named role bridged the gap. This asymmetry illustrates why governance must be institutionalized above the level of individual contracts: when oversight depends on informal arrangements between parties with misaligned incentives, accountability becomes structurally impossible.

\subsubsection{Traceability by default.} Stuxnet, a sophisticated computer worm, operated without traceability of any kind. Once deployed, it propagated autonomously across networks, executed sabotage against Iranian centrifuges, and concealed its effects by replaying recorded sensor data to plant operators~\cite{langner2011stuxnet}. No kill switch or recall mechanism was documented. Attribution required years of forensic analysis; accountability was effectively impossible in real time. The framework's requirement for tamper-evident logs with sensor provenance would have made scope creep detectable. In the Kargu-2 case, no traceability of autonomous engagement decisions was available to UN investigators~\cite{unpanel2021libya}. For Lavender, the reported absence of feature-level logging meant that the basis for individual targeting scores could not be reconstructed or challenged~\cite{abraham2024lavender}. In each case, the inability to reconstruct the decision chain rendered post hoc accountability procedurally impossible.

\subsubsection{Adversarial evaluation.} The Patriot system's 2003 friendly-fire incidents illustrate the cost of inadequate operational testing. A Defense Science Board investigation found that the system was given ``too much autonomy'' and that operators ``trusted the system in a naive manner''~\cite{dsb2005patriot}. The IFF subsystem misclassified friendly aircraft as hostile threats; negative testing (does the system correctly reject friendly aircraft under realistic confusion scenarios?) would have identified this failure mode before deployment. The USS Vincennes case prefigures the same concern at the interface level. The Aegis system's radar correctly tracked Iran Air Flight~655 as climbing in a civilian corridor, but the crew correlated a ground-based military IFF signal with the airborne contact~\cite{fogarty1988vincennes}. The system was never tested for this specific class of confusion. Adversarial evaluation under the proposed framework would require testing against precisely such scenarios: ambiguous IFF environments, overlapping military and civilian signatures, and high-stress time-critical conditions.

\subsubsection{The assistant principle.} Lavender exemplifies the failure of every requirement the assistant principle imposes. The system presented a name and a score without exposing the features driving the classification or any confidence interval. Human analysts reviewed each recommendation for approximately twenty seconds, a review that often consisted solely of verifying that the flagged individual was male~\cite{abraham2024lavender, hrw2024lavender}. This is performative endorsement. Calibrated uncertainty and forced abstention below confidence thresholds would have flagged the system's reported ten-percent error rate, corresponding to approximately 3,700 misidentifications among 37,000 flagged individuals. The Gospel system, which generated over 12,000 structural targets during operations in Gaza~\cite{hrw2024gaza, lieber2024gospel}, illustrates the same dynamic at the level of infrastructure: when machine-generated target volume overwhelms human review capacity, oversight becomes nominal. The Vincennes tragedy shows that interface design failures predate modern AI. The Aegis system displayed raw data without highlighting the anomaly between the aircraft's climbing trajectory and its putative hostile classification~\cite{fogarty1988vincennes}. Interface design that foregrounds uncertainty and anomalies, as the assistant principle requires, would have given the crew the perceptual salience needed to override.

\subsubsection{Technical enforceability.} Stuxnet's propagation beyond its intended target to approximately 115,000 systems in multiple countries, including allied nations~\cite{langner2011stuxnet}, is a concrete demonstration of what happens when hard scope limits are absent. The framework's Tier~5 requirements (target lists, network boundaries, time windows enforced at the code level) are designed to contain precisely this class of failure. The Kargu-2 operated without geographic, temporal, or target-class constraints enforced at the system level~\cite{unpanel2021libya}. Iron Dome again provides the positive case: its engagement criteria (ballistic trajectory prediction within defended zones) function as hard interlocks that bound the system's autonomous operation independently of software-level controls~\cite{cepa2024irondome}.

\begin{table}
    \caption{Cross-reference matrix: eight operational cases assessed against the six governance dimensions of the proposed framework. Every case involving documented harm exhibits deficiencies in at least two dimensions.}
    \label{tab:case_matrix}
    \centering
    \small
    \setlength{\tabcolsep}{4pt}
    \renewcommand{\arraystretch}{1.3}
    \begin{tabularx}{\textwidth}{l C C C C C C}
        \toprule
        \textbf{Case} &
        \textbf{Tiering} &
        \textbf{Account.} &
        \textbf{Trace.} &
        \textbf{Adv.\ eval.} &
        \textbf{Assist.} &
        \textbf{Enforce.} \\
        \midrule
        USS Vincennes & $\bullet$ & $\circ$ & $\star$ & $\circ$ & $\circ$ & $\bullet$ \\
        Patriot & $\bullet$ & $\star$ & $\star$ & $\circ$ & $\star$ & $\star$ \\
        Stuxnet & $\circ$ & $\circ$ & $\circ$ & $\star$ & - & $\circ$ \\
        Iron Dome & $\bullet$ & $\bullet$ & $\bullet$ & $\bullet$ & $\bullet$ & $\bullet$ \\
        Project Maven & $\circ$ & $\circ$ & $\star$ & $\star$ & $\star$ & $\star$ \\
        Kargu-2 & $\circ$ & $\circ$ & $\circ$ & $\circ$ & - & $\circ$ \\
        Gospel & $\star$ & $\star$ & $\star$ & $\star$ & $\circ$ & $\star$ \\
        Lavender & $\circ$ & $\star$ & $\circ$ & $\star$ & $\circ$ & $\star$ \\
        \bottomrule
    \end{tabularx}
    \footnotetext{$\bullet$ = dimension satisfied or not applicable; $\star$ = partially satisfied or ambiguous; $\circ$ = absent and absence contributed to documented harm; - = not applicable to this case.}
\end{table}

Table~\ref{tab:case_matrix} reveals a consistent pattern: every case involving documented harm exhibits deficiencies in at least two governance dimensions. No single safeguard would have been sufficient in isolation. The presence or absence of a human operator is orthogonal to governance quality; Lavender placed a human in the loop, while Iron Dome removed one, yet the latter satisfies the framework's requirements comprehensively. These are counterfactual analyses; the framework's predictive power can only be tested through prospective application in certification procedures. That limitation is acknowledged. What the analysis does establish is that each governance mechanism independently addresses a documented class of failure, and that their combination defines a governance posture substantially more demanding than any currently in force.
% ==========================================
%                 PART3
% ==========================================

% ==========================================
%                 CONCLUSION
% ==========================================
\section{Conclusion}
\label{sec:conclusion}

The governance of military AI is usually debated as a question of permission: what machines may be allowed to decide, and where the line of prohibition should fall. This paper has argued that a prior question determines whether any such line can be enforced. Systems whose reasoning cannot be inspected, whose evaluation can be gamed, and whose deployment disperses knowledge away from authority erode the conditions under which human judgment remains judgment at all. The presence of an operator guarantees nothing about the quality of oversight; what matters is whether that operator retains the temporal authority to intervene and the interpretive access to know when intervention is warranted.

Taking this diagnosis seriously requires abandoning a tempting ambition. If ethical constraints cannot be reliably encoded in a loss function, they must be encoded in institutions: named roles that make responsibility attributable, evaluation regimes that resist optimization, deployment thresholds calibrated to irreversibility, and interfaces that keep uncertainty visible. This is the practical form of Crootof's move from intent to procedural integrity. It does not answer the philosophical question of who bears moral responsibility for an algorithmic output, and it is not meant to. It substitutes a question that institutions can actually adjudicate: what process was followed, by whom, and did it meet a documented standard? The counterfactual analysis in Sect.~\ref{ssec:operational_precedents} indicates that this substitution is not merely conceptual, since the failures it maps are procedural failures rather than failures of intention.

The proposal has clear limits. The tiered classification is schematic, and operationalizing it would demand sustained negotiation among states with divergent legal traditions and procurement cultures; the history of defense standardization suggests such convergence is incremental at best. The hybrid trajectory presumes continued progress in interpretability and uncertainty quantification at a pace no one can guarantee. Most fundamentally, the framework addresses state military organizations operating within treaty regimes, and offers little purchase on non-state actors who recognize no such obligations.

These limits mark the work that follows. Structured expert consultation, for instance through a Delphi process involving military legal advisers, operational commanders, and engineers, would test whether the tiering and the accountability roles survive contact with institutional practice. Section~\ref{ssec:existing_instruments} points out where current rules stop. Rather than inventing new principles, this framework simply focuses on how to apply existing ones to the AI models and military contexts that are currently left out. And the interface principles of Sect.~\ref{ssec:assistant_principle} invite direct empirical study: whether forced abstention and salient uncertainty measurably improve decision quality under time pressure is a behavioral question, and one the framework currently answers only by assertion.

The deliberations of the UN Convention on Certain Conventional Weapons, now approaching a possible Review Conference, will test whether states are prepared to convert declaratory principles into binding operational constraints. Whatever form those constraints take, the assessment of whether a particular system, in a particular context, satisfies the demands of international humanitarian law will remain a human judgment. The task is to ensure that it is exercised with adequate knowledge and real authority, rather than performed after the fact by someone with neither.
% ==========================================
%                 CONCLUSION
% ==========================================

\section*{Declarations}

\subsection*{Funding}
No funding was received for conducting this study.

\subsection*{Competing interests}
Authors A, B and D are employed by SAS Impact, which uses machine learning systems for defense applications. Author C declares no competing interests. The authors received no specific funding for this work.

\subsection*{Data availability}
No datasets were generated or analyzed during the current study. All cases discussed are documented in publicly available sources cited in the reference list.

\bibliography{main}
\end{document}